\documentclass[12pt]{iopart}
\usepackage{iopams} 
\usepackage{graphicx}

\begin{document}

\title[Inter-dot coupling and excitation transfer of InAs QDs]
{Inter-dot coupling and excitation transfer mechanisms of telecommunication band InAs quantum dots at elevated temperatures}

\author{C~Hermannst\"{a}dter$^1$, N~A~Jahan$^{1,2}$, J-H~Huh$^1$, H~Sasakura$^1$, K~Akahane$^3$, M~Sasaki$^3$ and I~Suemune$^1$}

\address{$^1$ Research Institute for Electronic Science, Hokkaido University, Sapporo~001-0021, Japan}
\address{$^2$ Graduate School of Information Science Technology, Hokkaido University, Sapporo~060-0814, Japan}
\address{$^3$ National Institute of Information and Communications Technology, Koganei, Tokyo 184-8795, Japan}

\ead{\mailto{claus@es.hokudai.ac.jp}}


\begin{abstract}

We investigate the photoluminescence temperature dependence of individual InAs/InGaAlAs quantum dots emitting in the optical telecommunication bands. The high-density dots are grown on InP substrates and the selection of a smaller dot number is done by the processing of suitable nanometer sized mesas. Using ensembles of only a few dots inside such mesas, their temperature stability, inter-dot charge transfer, as well as, carrier capture and escape mechanisms out of the dots are investigated systematically. This includes the discussion of the dot ensemble and individual dots. Among the single-dot properties, we investigate the transition of emission lines from zero-phonon line to acoustic phonon sideband dominated line shape with temperature. Moreover, the presence of single recombination lines up to temperatures of around 150~K is demonstrated.

\end{abstract}


\vspace{2pc}
\maketitle

\small
\tableofcontents
\vspace{1pc}
\normalsize

\markboth{Inter-dot coupling and excitation transfer of InAs QDs}{Inter-dot coupling and excitation transfer of InAs QDs}

\section{Introduction}
\label{sec:Introduction}

Solid state semiconductor single photon emitters, such as, self-assembled quantum dots (QDs) \cite{1a,1b,1c}, have recently attracted plenteous attention for a wide range of application in quantum information and communication \cite{2a,2b,2c,2d,2e}. They can be used as deterministic photon sources, comparable to single atoms and molecules, impurities in semiconductors, etc. 
Semiconductor QDs are powerful candidates for single photon and entangled photon pair generation for applications in state-of-the-art highly secure communication and quantum computation \cite{3a,3b}. An operating range in the telecommunication O and C bands (1.3 and 1.55 $\mu$m) is favourable to insure the applicability of standard silca based fibres and fibre networks. High photon extraction efficiency from single QD devices is another significant factor to suppress the bit error rate. 
The inter-dot coupling \cite{Zho,q1,q2,q3,q4,q5,q6} as well as the interaction of the QDs and the confined carriers with their environment are important phenomena, either to exploit them to manipulate the quantum state \cite{i1,i2}, or to suppress them in order to keep the QD of interest as isolated as desired. A detailed understanding and description of these interactions gains importance when the system temperature is increased above liquid Helium temperature (4.2 K) and towards technically easier achievable temperatures above liquid Nitrogen temperature (77 K), because thermally activated processes come into play.

We realized metal-embedded tapered mesa (nano-cone) structures to meet the challenge of studying a low number of QDs in the telecommunication O and C bands and to provide low loss photon extraction, in a comparable approach to the demonstration by Takemoto, et al.~\cite{Tak1}. The advantages of the proposed and realized nano-cones are their fast and highly reproducible nanometer- (or micrometer-)sized fabrication, the possibility to integrate them in more complex devices and the efficient light extraction by embedding them inside reflective metal. 
In this paper we briefly discuss the nano-cone structure fabrication process with the main purpose of isolating individual dots and creating nanostructures of efficient photon extraction. 
The excitons confined to individual QDs as well as their relaxation and recombination are investigated in detail with the focus on their temperature dependence. The thermally induced exciton transfer between and escape from QDs is described using a simple model. The optical spectroscopy results furthermore present wavelength-tunable/selectable single QD emission towards the realization of a practical single photon source in both the telecommunication O and C bands.

\section{Samples and experiment}
\label{sec:Experiment}

\begin{figure}[b]
	\centering
	\includegraphics[scale=.25]{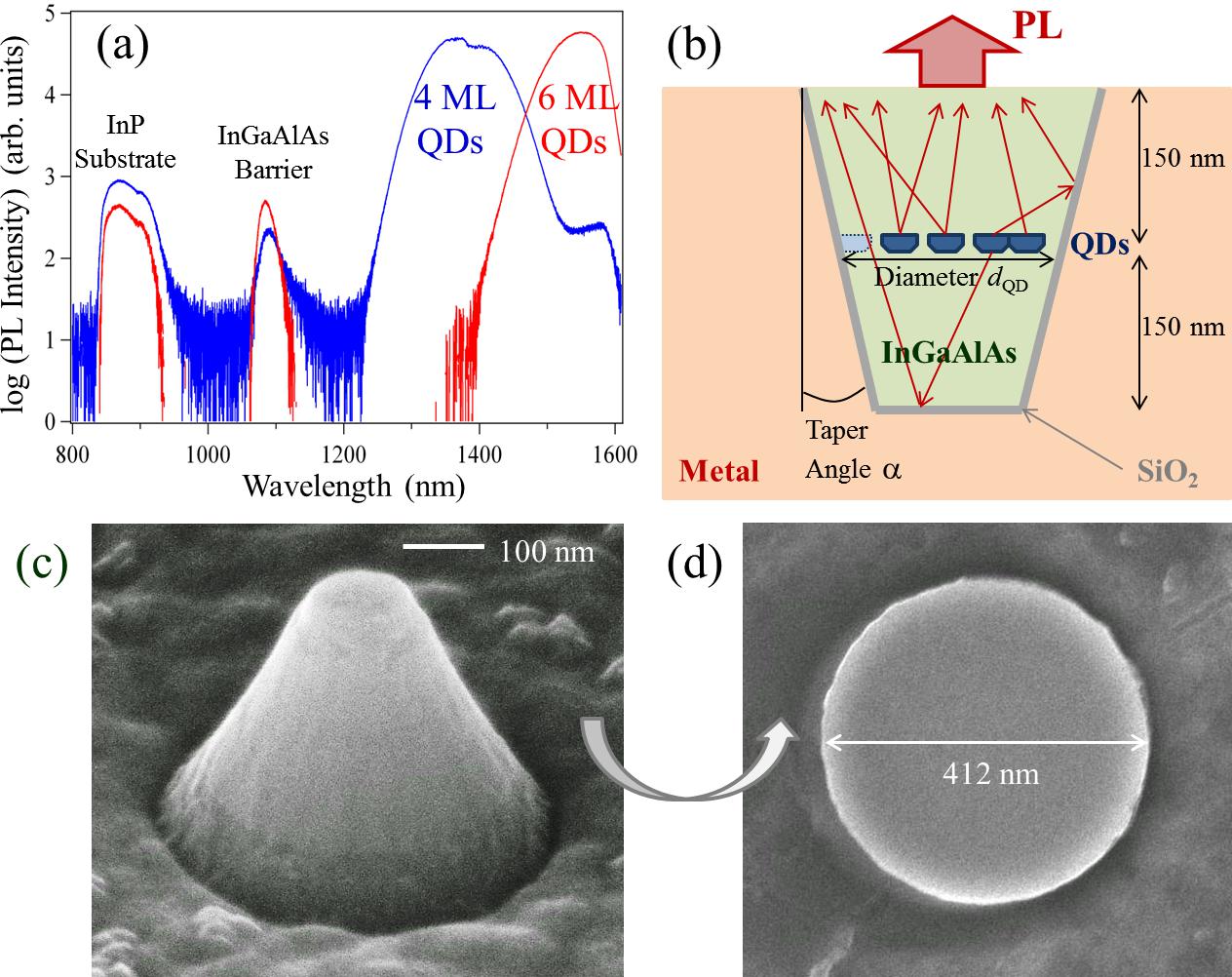}
	\caption{(a) PL of the 4 and 6 ML InAs QD samples at 10 K, including the emission from the InP substrate and the In$_{0.53}$Ga$_{0.22}$Al$_{0.25}$As barrier. (b) Schematic illustration of the PL structure, a metal-embedded cone, with a thin insulator layer (SiO$_2$/Si$_3$N$_4$) between the metal and the semiconductor part and the QDs located in the center of the InGaAlAs barrier. (c) Side view SEM image of an as etched 26$^\circ$ tapered nano-cone with a tip diameter of around 100 nm and a diameter of the QD containing plane of $d_{\rm QD}\approx250$~nm, (d) top-view SEM image of the respective turned-around and SiO$_2$/Ti-embedded nano-cone with removed substrate.}
	\label{fig:Fig1}
\end{figure}

The investigated samples contain high-density ($\sim 1 \times 10^{11}$ cm$^{-2}$) InAs QDs grown on a lattice matched 150 nm thick In$_{0.53}$Ga$_{0.22}$Al$_{0.25}$As buffer layer on InP(311)B substrates by molecular beam epitaxy \cite{growth,Aka1}. For optical measurements the QDs were capped  with another 150 nm thick In$_{0.53}$Ga$_{0.22}$Al$_{0.25}$As barrier layer. QD emission between 1.2 and 1.6~$\mu$m ($1.03-0.77$ eV) could be achieved by varying the nominal thickness of the optically active QD layer between four and six monolayers (MLs); the photoluminescence (PL) of these as grown, unstructured samples is displayed in \fref{fig:Fig1}~(a). These QD samples have no wetting layer, which can be concluded from the PL spectra, where no indication for the respective luminescence is found \cite{xx}. The absence of a wetting layer has vast consequences on the QD optical properties and charge redistribution processes, which will be discussed in the following sections.
Sequential steps of electron-beam lithography and etching allowed for a well-controlled and high yield fabrication of nano-cones with typical 26$^{\circ}$ taper angles, heights of 300~nm and tip diameters down to less than 100~nm (\fref{fig:Fig1}~(b, c)). After the deposition of a thin insulating layer ($10-60$ nm SiO$_2$ or Si$_3$N$_4$), the nano-cones were embedded in metal (Titanium (Ti) or Silver (Ag)). In the subsequent step, the InP substrate was removed and the sample was turned upside down, which is highlighted in the schematic illustration and scanning electron microscope (SEM) top view image in \fref{fig:Fig1}~(b) and (d) (more details on the fabrication are presented in Huh, et al.~\cite{Huh1}). In the further text the diameter of the QD containing plane, $d_{\rm QD}$, of the nano-cones is given when nano-cone sizes are specified.

The PL of such turned-around and metal-embedded nano-cones was investigated in a micro-PL setup optimized for the near infrared (NIR) spectral range. The samples were cooled in a Helium-flow cryostat to access a temperature range from 4 K to room temperature. For non-resonant excitation a HeNe laser (632.8 nm; 1.959 eV) was focused on the sample using an NIR-coated $\rm NA=0.42$ objective lens, which results in a spot size of approximately 1.5 $\mu$m. The PL was collected with the same objective lens and dispersed by a 50 cm double monochromator with 300 gr./mm grating and detected by a liquid Nitrogen cooled InGaAs-photodiode array detector (cutoff $\approx$ 1.6 $\mu$m).

\section{Quantum dot luminescence at around 1.3 $\mu$m}
\label{sec:1.3}

In this section the focus of the experimental results and the following discussion is based on the smaller 4 ML QDs which have their low temperature PL centered at around 1.3 $\mu$m (0.95 eV). Using these QDs the temperature dependence up to the complete quenching can be investigated within the sensitivity range of our detector. The PL wavelength furthermore coincides with the telecommunication O band.

\subsection{Quantum dot photoluminescence power and temperature dependence}
\label{sec:QDPLPowerAndTemperatureDependence}

\begin{figure}
	\centering
	\includegraphics[scale=.5]{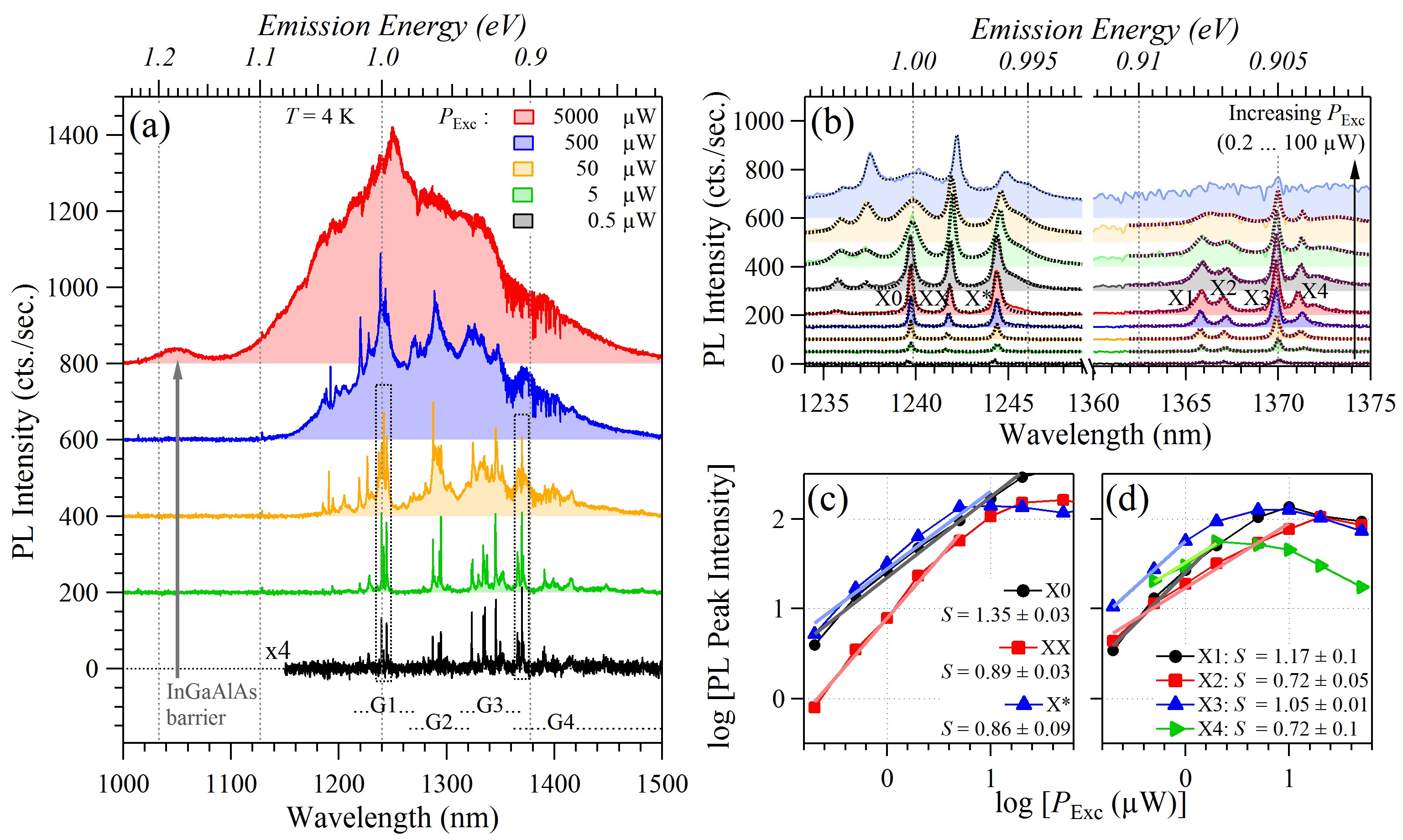}
	\caption{(a) PL power dependence of a $d_{\rm QD}\approx250$~nm SiO$_2$/Ti-embedded nano-cone under non-resonant excitation at 4 K. Four groups (G1--G4) of QD emission lines are indicated according to their spectral position for later identification. (b, c) Three PL lines of one single QD at around 1 eV and (b, d) four PL lines at around 0.905 eV at different excitation powers. The power-dependent integrated PL peak intensities are fitted with a power law ($I \propto P_{\rm Exc}^S$).}
	\label{fig:AaPowerAll}
\end{figure}

The complete PL spectrum from a nano-cone ($d_{\rm QD}\approx250$~nm) is displayed in \fref{fig:AaPowerAll}~(a), where at low excitation power a total of around ten optically active QDs are visible between around $0.88$ and $1.02$ eV. With increasing excitation power additional PL lines at higher energy emerge which are due to excited state recombination in the QDs. At the highest excitation power, the InGaAlAs barrier luminescence can be seen at around $1.180$ eV; no wetting layer exists for this type of QD samples. 
For the first 4 ML QD, which is shown on the left side of the more detailed close-up \fref{fig:AaPowerAll}~(b, c), X0 and X* exhibit an almost linear dependence ($S \approx 0.9$) of their intensities, $I$, on excitation power, $P_{\rm Exc}$, in the respective power law, $I \propto P_{\rm Exc}^S$. This indicates their origin from neutral and (negatively) charged exciton recombination. XX shows a super-linear dependence ($S \approx 1.35$) indicating its origin from biexciton recombination.
The second presented dot on the right side of the close-up, \fref{fig:AaPowerAll}~(b, d), does not show the same clear power dependence; however, X1 and X3 depend slightly super-linearly on the excitation power, whereas X2 and X4 exhibit a sub-linear power dependence. Given the relative peak energies and the power dependencies, X1 ($S \approx 1.17$) might be due to a charged biexciton or a positive trion and X3 ($S \approx 1.05$) due to a neutral biexciton. The other two peaks are likely assigned as follows: X2 ($S \approx 0.72$) to a neutral exciton and X4 ($S \approx 0.72$) to a (negative) trion. For the following discussion the exact origin of the recombination lines is not further discussed and of no major importance.

\begin{figure}
	\centering
	\includegraphics[scale=.5]{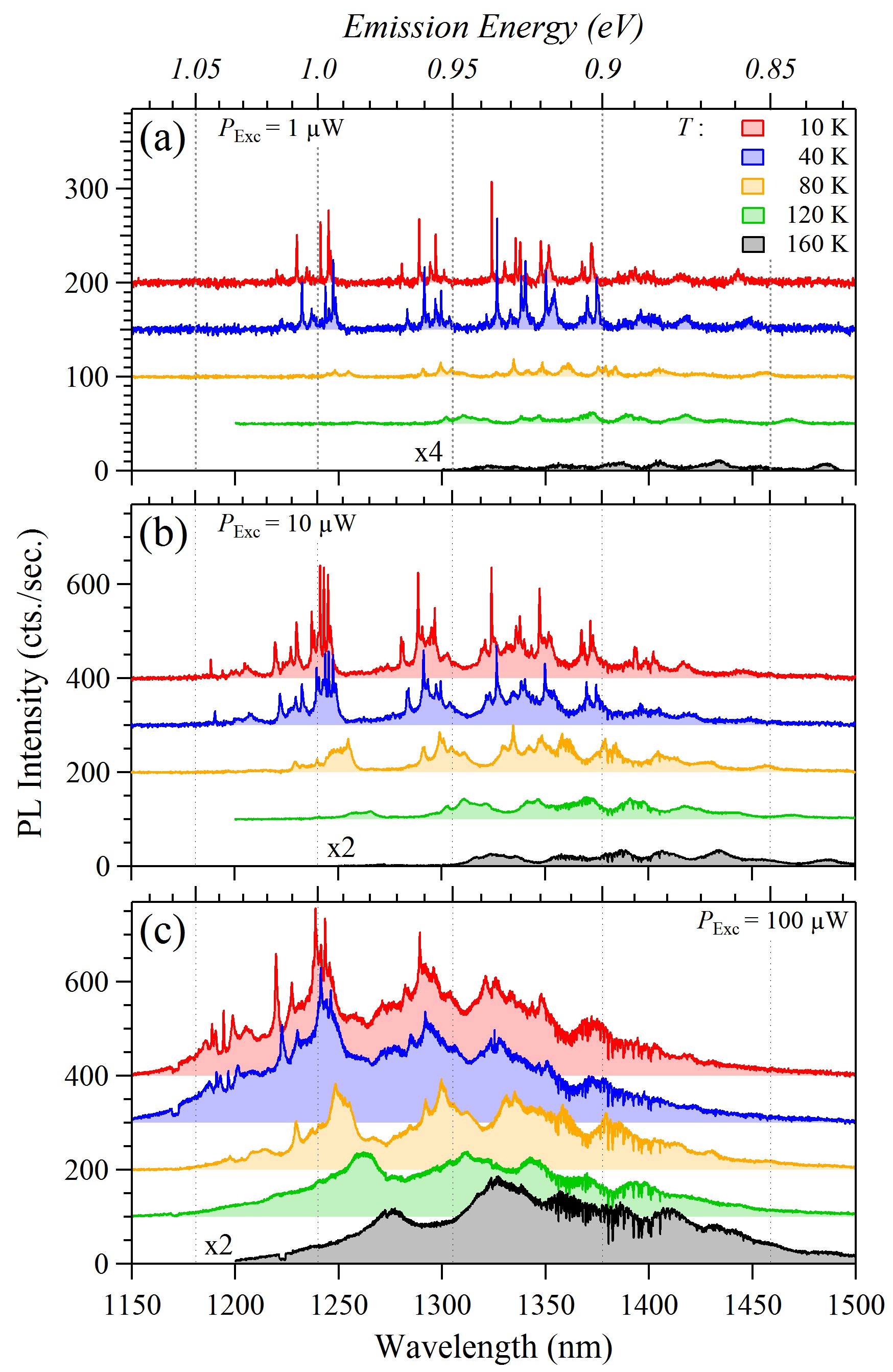}
	\caption{PL temperature dependence from the same nano-cone shown in \fref{fig:AaPowerAll} between 10 and 160 K for three different excitation conditions: (a) low power (1 $\mu$W, well below exciton saturation), (b) medium power (10 $\mu$W, approaching saturation), and (c) high power (100 $\mu$W, above saturation) non-resonant excitation.}
	\label{fig:AaTemp}
\end{figure}

The PL of the same nano-cone with all contained optically active QDs was examined depending on the excitation power and as function of the sample temperature between 10 and 160 K. The spectra obtained under low, medium and high (above exciton saturation) power conditions are displayed in \fref{fig:AaTemp}. All three presented data sets, (a--c), are to compare, i.e., their scalings and relative offsets are chosen identically, only the lowest-power data (a) are displayed on half the intensity scale and thus with half the offsets for more clarity.
Comparing the PL temperature dependence for the different excitation regimes the first obvious observation is that for higher power excitation the PL quenching seems much less pronounced, i.e., the PL persists up to higher temperatures. This might be intuitively attributed to refilling processes that take place under non-resonant excitation, where the carriers can repopulate the QDs which leads to a ``slower" quenching of all dots. 
Another rather obvious qualitative observation is the emergence of the excited states which results in a major modification of the PL temperature dependence in the high-power case (\fref{fig:AaTemp}~(c)).
More detailed examination, analysis and discussion of the temperature dependence of both the complete nano-cone PL (QD ensemble) and of individual QD PL (single QDs) is carried out in the following section. 

\subsection{Analysis and discussion}
\label{sec:Disc1.3}

The temperature dependent PL is analyzed by fitting with a model that accounts for both the carrier capture and the quenching temperature dependence,


\begin{equation}
	I(T) = I_0\left( p + \frac{1-p}{\exp{\left[k_{\rm B} T_0 / (k_{\rm B} T)\right]}}\right)	\times \frac{1}{ 1 + \sum\limits_{i=1,2} b_i \exp{\left[-E^{({\rm a}i)} / (k_{\rm B} T)\right]}}\ .
\label{Eq:Arrha2}
\end{equation}

Therein, the first factor describes the temperature dependent excitation (carrier capture) which takes place under non-resonant excitation and the second factor describes the thermal quenching of the states for increasing temperatures \cite{Gel}, which includes two distinct escape mechanisms labeled with the index $i=1,2$. The parameters in the excitation term are the total (maximum) intensity $I_0$, the dimensionless weighting parameter $p$, which defines the ratio between temperature independent and temperature dependent excitation contributions, and the temperature constant $T_0$, which defines a temperature equivalent for the excitation process; $k_{\rm B}$ is the Boltzmann constant and $T$ is the system temperature. In the quenching term, $E^{({\rm a}i)}$ are activation energies that are assigned to a certain escape mechanism and $b_i$ are the corresponding dimensionless fitting coefficients. The meaning of these coefficients $b_i$ can be understood as the ratio of escape and capture rates of the described mechanism, i.e., $b_i\propto$ (escape-rate)$_i/$(capture-rate)$_i$, as it is discussed and derived in a comprehensive fashion by Le Ru, et al. in~\cite{LeR}. 

\begin{figure}
	\centering	\includegraphics[scale=.5]{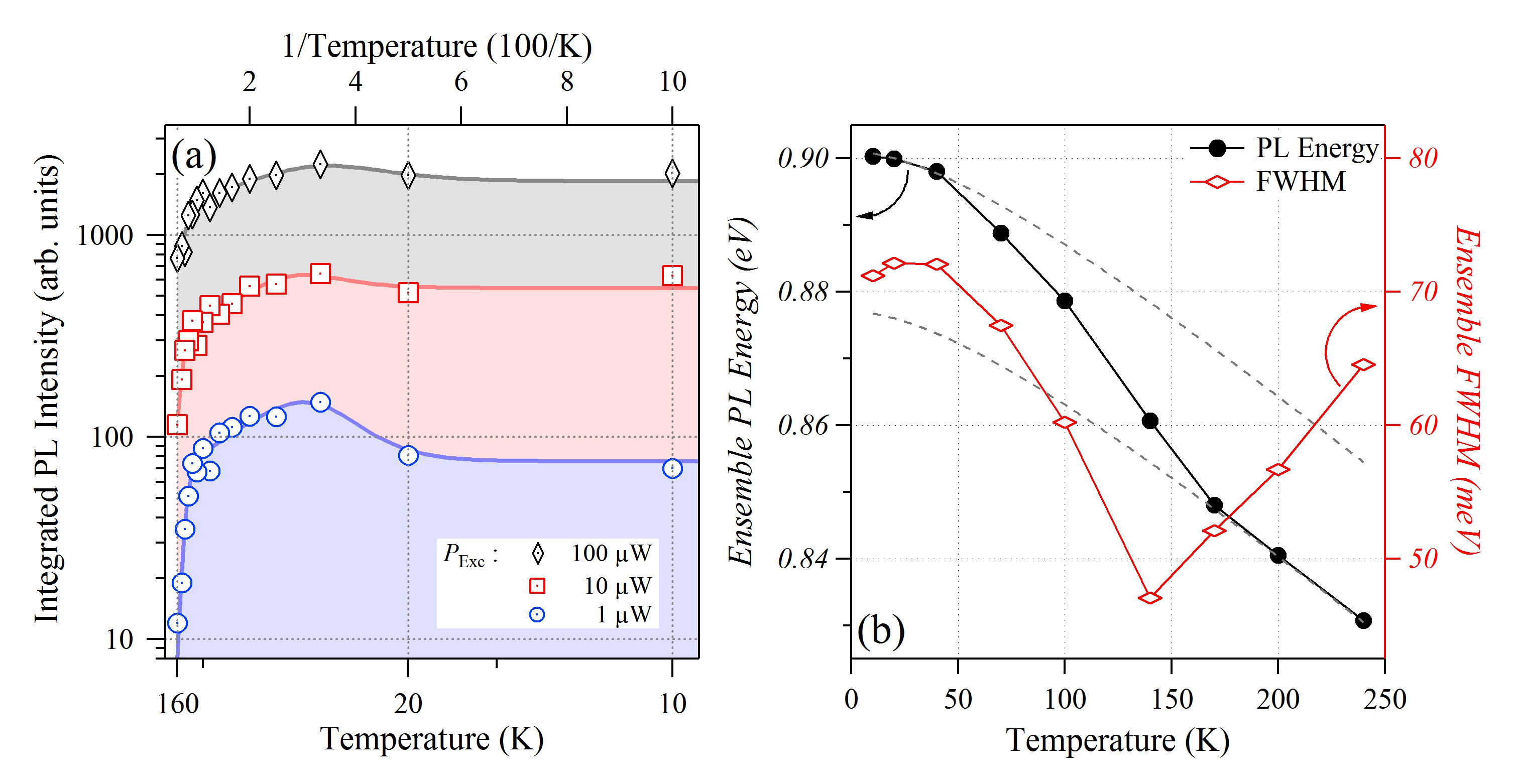}
	\caption{(a) Temperature dependence of the integrated PL of the complete spectra displayed in \fref{fig:AaTemp} for low, medium and high excitation powers (markers) and the respective fits with equation \eref{Eq:Arrha2} (filled solid lines). (b) Temperature dependence of the ensemble PL data acquired from an unstructured QD sample for comparison \cite{xx}; the emission energy of the ensemble (black circles, obtain from Gaussian line fits), with the dashed grey guide lines obtained from fitting the low and high temperature data by Varshni empirical equation ($E(T) = E_0 - (a/T^2)/(b + T)$), and the ensemble PL FWHM (red diamonds).}
	\label{fig:AXEns}
\end{figure}
\begin{table}
	\centering \footnotesize
		\begin{tabular}{c|*{3}c|*{4}c} \hline 
			$P_{\rm Exc}$ & $I_0$ & $p$ & $T_0$ \footnotesize (K) & $b_1$ & $E^{({\rm a}1)}$\footnotesize (meV) & $b_2$ & $E^{({\rm a}2)}$\footnotesize (meV) \\ \hline
			100 $\mu$W & $2.2\cdot10^5$ & $0.0084$ & $129\pm 1$ & $184\pm1$ & $13.0\pm0.1$ & $1.4\cdot10^4\pm0.02\cdot10^4$ & $\ 73\pm1$ \\
			10 $\mu$W & $1.0\cdot10^5$ & $0.0055$ & $175\pm3$ & $410\pm6$ & $18.4\pm0.3$ & $3.6\cdot10^6\pm0.2\cdot10^6$ & $135\pm6$ \\
			1 $\mu$W& $5.5\cdot10^4$ & $0.0014$ & $166\pm6$ & $1058\pm73$ & $18.0\pm0.8$ & $5.4\cdot10^7\pm1.1\cdot10^7$ & $139\pm23$ \\ \hline 
		\end{tabular}
	\caption{Fit parameters (equation \eref{Eq:Arrha2}) for the integrated PL from the nano-cone, i.e., from the contained QD sub-ensemble; the $1\sigma$-fit-errors are omitted when of no significant magnitude ($<<1\%$ of the parameter value).}
	\label{tab:IntegratedPL}
\end{table}

We start with a comprehensive analysis of the ensemble PL of all QDs within the investigated nano-cone, which is later referred to as QD sub-ensemble. This is done by integration of the total PL intensities within the spectral range displayed in \fref{fig:AaTemp}. The obtained integrated intensities for the three different excitation powers are displayed as a function of inverse temperature in \fref{fig:AXEns}~(a). The selection of displaying the inverse temperature dependence has been made to show the low temperature side in a more distinctive way. This addresses the fact that especially the carrier capture (and also re-excitation) processes are heavily dependent on the power regime. The experimental data were fitted using the afore introduced equation \eref{Eq:Arrha2} and all obtained fitting parameters are summarized in table \ref{tab:IntegratedPL}. 

On the excitation side, it is obvious that in case the excitation power remains below the exciton saturation, i.e., for 1 $\mu$W (blue circles) and 10 $\mu$W (red squares), the carrier capture process is rather inefficient, which is revealed by the large obtained characteristic temperature $T_0\approx 170$ K and the small values of the corresponding constant $p\approx0.0014$ and 0.0055. The increase in PL intensity up to 30 K in the respective data-sets highlights this behavior.
For increasing excitation power, the carrier capture process becomes more efficient or, in other words, less temperature sensitive, indicated by increasing values of $p$ and decreasing values of $T_0$. This trend is highlighted by the flattening of the data on the low temperature side with increasing excitation power. For high power excitation above the exciton saturation (100 $\mu$W, black diamonds), also $T_0$ significantly decreases to 129 K and $p$ further increases.

On the decay side, the following trend can be seen. For both the low and medium power excitation regimes, which are with no significant contribution of excited states, very similar behavior is observed. This includes the two different escape mechanisms with activation energies of $E^{({\rm a}1)}\approx18$ meV and $E^{({\rm a}2)}\approx140$ meV, respectively, which are almost the same for the low and medium power conditions. For the high power case, in contrast, especially the larger activation energy becomes significantly smaller, $E^{({\rm a}2)}\approx73$ meV. 
We assign the first mechanism, $i=1$, to a process unique for high-density QD ensembles, which is related to inter-dot coupling \cite{Zho}, and the second mechanism, $i=2$, to the thermal carrier escape from QDs to the barrier. 

The discussion of the second mechanism, which leads to the significantly reduced activation energy at high excitation power, directly explains the impact of the populated excited states and can be understood rather straight forwardly and intuitively. The carrier escape is directly related to non-radiative carrier escape to the InGaAlAs barrier, which embeds the QDs, and leads to a linear relation between the barrier-dot energy separation and the activation energy $E^{({\rm a}2)}$. The energy separation between the barrier and the QD excited states (no matter what kind of states they are) is significantly lower than for the QD ground states, and thus their enhanced population at higher excitation rates leads to a reduction of the corresponding activation energy $E^{({\rm a}2)}$. 
The accompanying fact that also $b_2$ is smallest at high excitation power supports the aforementioned occurrence of re-excitation, as it indicates a larger corresponding carrier (re-)capture-rate. More details, including a quantitative discussion and explanation of the activation energy magnitudes, is carried out for single QDs in section \ref{sec:RelativeQDIntensitiesCouplingAndQuenching}.

The first escape mechanism, however, needs a more in-depth description, which is provided by taking a closer look at an ensemble of the same type of QDs (\fref{fig:AXEns} (b)). For this reference measurement, an unstructured sample and a macroscopic PL setup was used in order to simultaneously access a larger number of QDs for a better statistical distribution and thus ensemble average \cite{xx}. The temperature dependence of a QD ensemble under low power excitation conditions indicates that, first, the peak emission energy exhibits a pronounced shift to lower energies when compared to the expected trend using a fit by Varshni empirical equation and, second, the PL ensemble line width decreases with increasing temperature up to around 140 K. These observations can be well explained by a charge-carrier redistribution from the higher to the lower energy dots of the ensemble, which is promoted by an increased temperature and thus enhanced phonon scattering and thermal excitation. 
In order to conclude a corresponding carrier escape process from these rather obvious facts, we suggest to introduce coupled excited states (CES) which are delocalized over many (or even all) dots of the sub-ensemble. When carriers are ones excited to such a delocalized state, they can either relax favourably to the lowest energy states, which results in the redistribution in favour of the lower energy dots, or be subsequently further excited to the barrier and thus escape the QDs. This process is also discussed in more detail for individual QDs in section \ref{sec:RelativeQDIntensitiesCouplingAndQuenching}, where the relative change of intensities and the change of related activation energies allows for much better insight.
To apply this mechanism on the herein described QD sub-ensemble, it shall only be added, that for all excitation conditions such small activation energies to a CES could be obtained ($E^{({\rm a}1)}\approx13-18$ meV). Moreover, with increasing power, the corresponding $b_1$ parameter decreased, which again indicates an enhancement of the corresponding (re-)capture-rate, i.e., the relaxation from CES to a QD ground state.

We want to add that in contrast to typical self-assembled InAs/GaAs QDs, a remarkably similar behavior for ensembles of high-density InAs/GaAs quantum posts has been reported \cite{He,Kre,Vol}. This includes both the intensity increase and the charge redistribution from higher to lower energy states with increasing temperature. In analogy to our QDs discussed in the present paper, also these quantum posts are not coupled to a wetting layer rather than to a surrounding matrix-quantum well as illustrated in \cite{Vol}.

\subsubsection{Single quantum dot spectral properties.}
\label{sec:SingleQDSpectralChange}

\begin{figure}
	\centering
	\includegraphics[scale=.5]{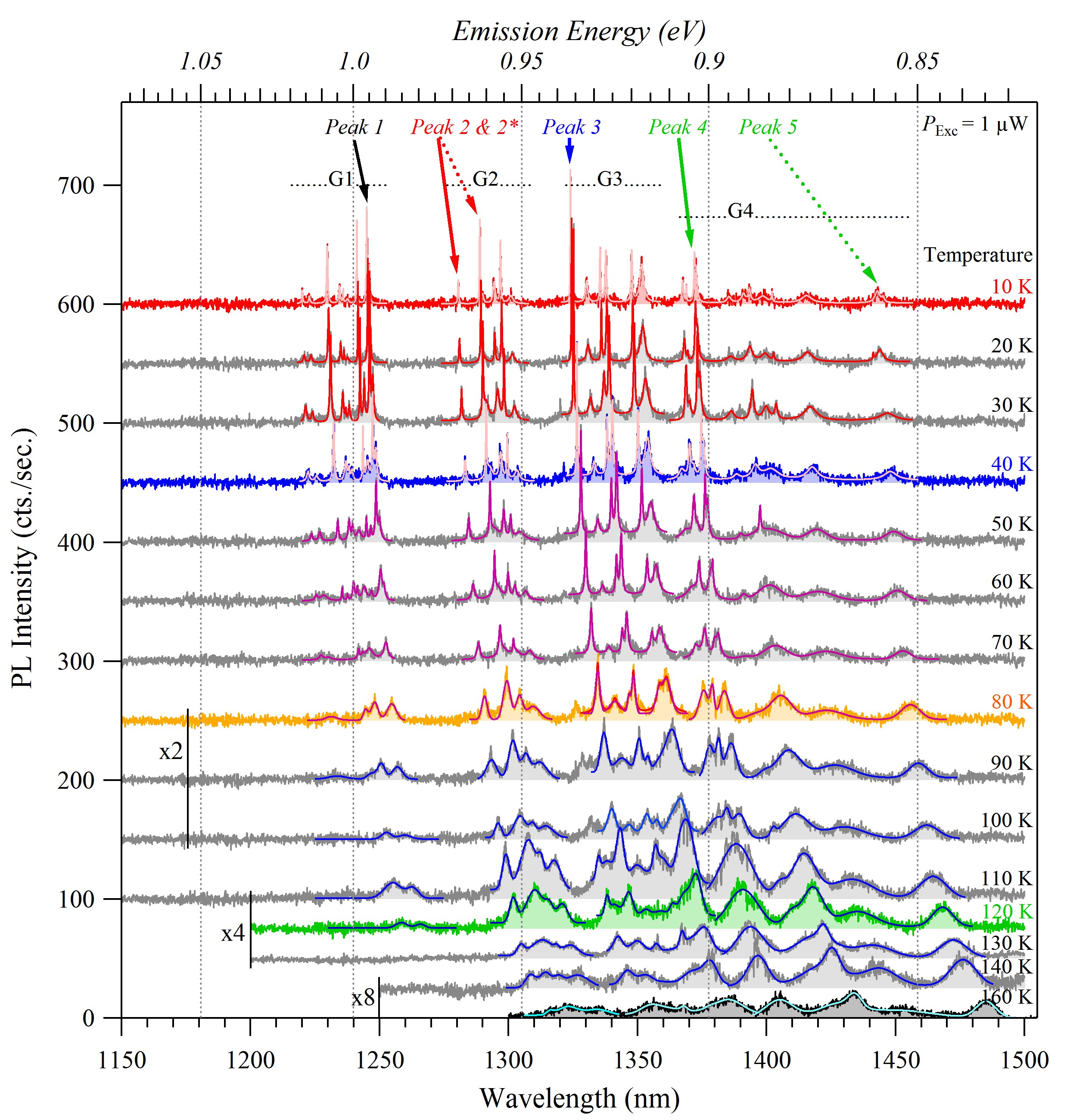}
	\caption{Detailed temperature series ($10-160$ K) under low power non-resonant excitation and line fits to all QD spectra. The PL lines, \emph{``Peaks 1--5''} are highlighted for later identification. The data and respective line colours at 10, 40, 80, 120 and 160 K are the same as shown in \fref{fig:AaTemp}~(a). The line fitting was obtained using quasi-Voigt functions and using a superposition of Lorentzian and Gaussian functions with changing weight for increasing temperatures (details in \fref{fig:AbPeak3}).}
	\label{fig:AaT001All}
\end{figure}

In the next step we investigate the temperature dependent phenomena within the nano-cone in a more detailed way and from a microscopic view point with the purpose to describe what happens on the single QD level and what promotes the inter-dot coupling. Therefore, we take another and closer look on the temperature dependence of the nano-cone PL under low excitation power, which is represented by the detailed data set displayed in \fref{fig:AaT001All}. This in-depth analysis includes multi-peak fits to all featured QD PL lines, accounting for changing line shapes and emission energies with increasing temperature.

\begin{figure}
	\centering
	\includegraphics[scale=.5]{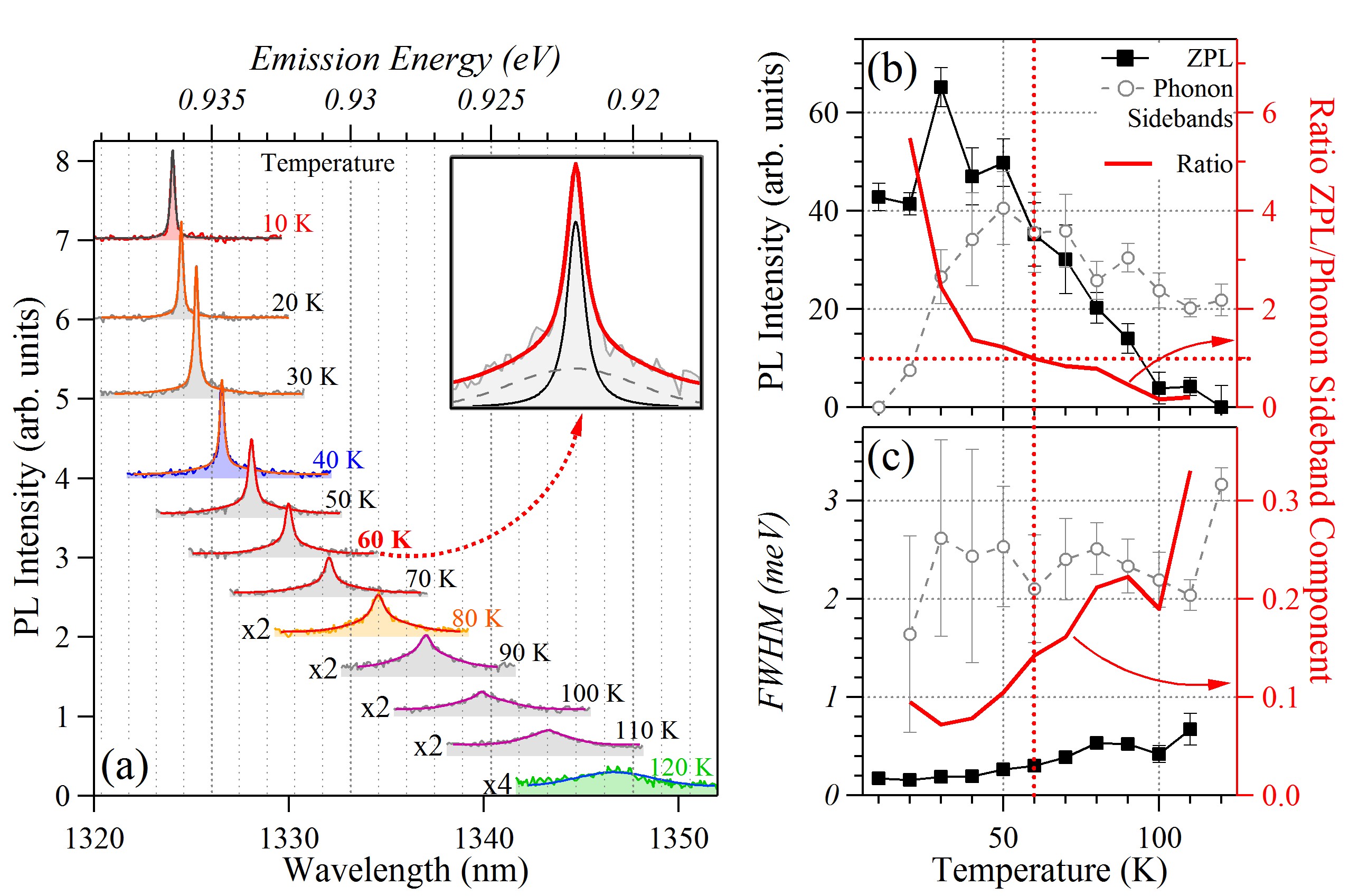}
	\caption{(a) Zoomed view of \emph{``Peak 3''} from \fref{fig:AaT001All} to determine the temperature dependent line shape transition from the pure ZPL to the predominantly acoustic phonon broadened spectral appearance, applying Lorentzian and Gaussian fits, respectively. 
	The inset presents the fit curves to the 60 K data, i.e., the total fit (thick red line), the ZPL part (solid black line) and the phonon sideband part (dashed grey line).
	(b) Temperature dependence of the ZPL (Lorentzian, black squares) and phonon sideband (Gaussian, grey circles) components of the integrated PL intensity and (c) the respective line width (FWHM) components; the red lines indicate the ratios of the ZPL to phonon sideband values.}
	\label{fig:AbPeak3}
\end{figure}

The behaviour of one single PL line, \emph{``Peak 3''}, is exemplified in \fref{fig:AbPeak3}. The zero-phonon line (ZPL), which is represented by the Lorentzian component with an initial line width of 0.17 meV, dominates the spectral shape at low temperatures and initially gains intensity for T $\leq$ 30 K and then starts to fade for T $>$ 30 K. At around $50-70$ K the acoustic phonon sidebands, represented by the Gaussian component, exhibit comparable intensities \cite{Ari}. Additional to acoustic phonon broadening, also charge fluctuations and contributions of additional capture and decay processes related to QD coupling and higher energy phonons might contribute to this broadening for increasing temperatures. The transition regime is highlighted for 60 K as inset in \fref{fig:AbPeak3}~(a), featuring the two distinct components, the ZPL (solid black line) and the phonon sidebands (dashed grey line). For T $>$ 80 K the phonon sideband intensity decreases slowly, while the ZPL rapidly diminishes and eventually vanishes completely at around 120 K (\fref{fig:AbPeak3} (b)). This particular recombination line can at least be identified as a single PL peak up to around 100 K without significant background or overlap from other features. 
Although it remains visible for temperatures up to 160 K, eventually background and overlapping of different features become more significant for T $\geq$ 120 K (\fref{fig:AaT001All}).

The initial intensity increase up to 30 K can be explained as before by a not perfectly efficient carrier capture from the barrier to the QD under low power non-resonant excitation. The temperature dependence of the initial carrier capture is again included in the quantitative discussion and modeling of the individual QD temperature dependence, which follows in section \ref{sec:RelativeQDIntensitiesCouplingAndQuenching}.
The relative total line width broadening can be assigned to the two main contributions that are highlighted in \fref{fig:AbPeak3} (c), the broadening of the ZPL, which will be discussed later, and the change of relative weight between the narrower ZPL and the broader acoustic phonon sidebands with increasing temperature. 

\begin{figure}
	\centering
	\includegraphics[scale=.5]{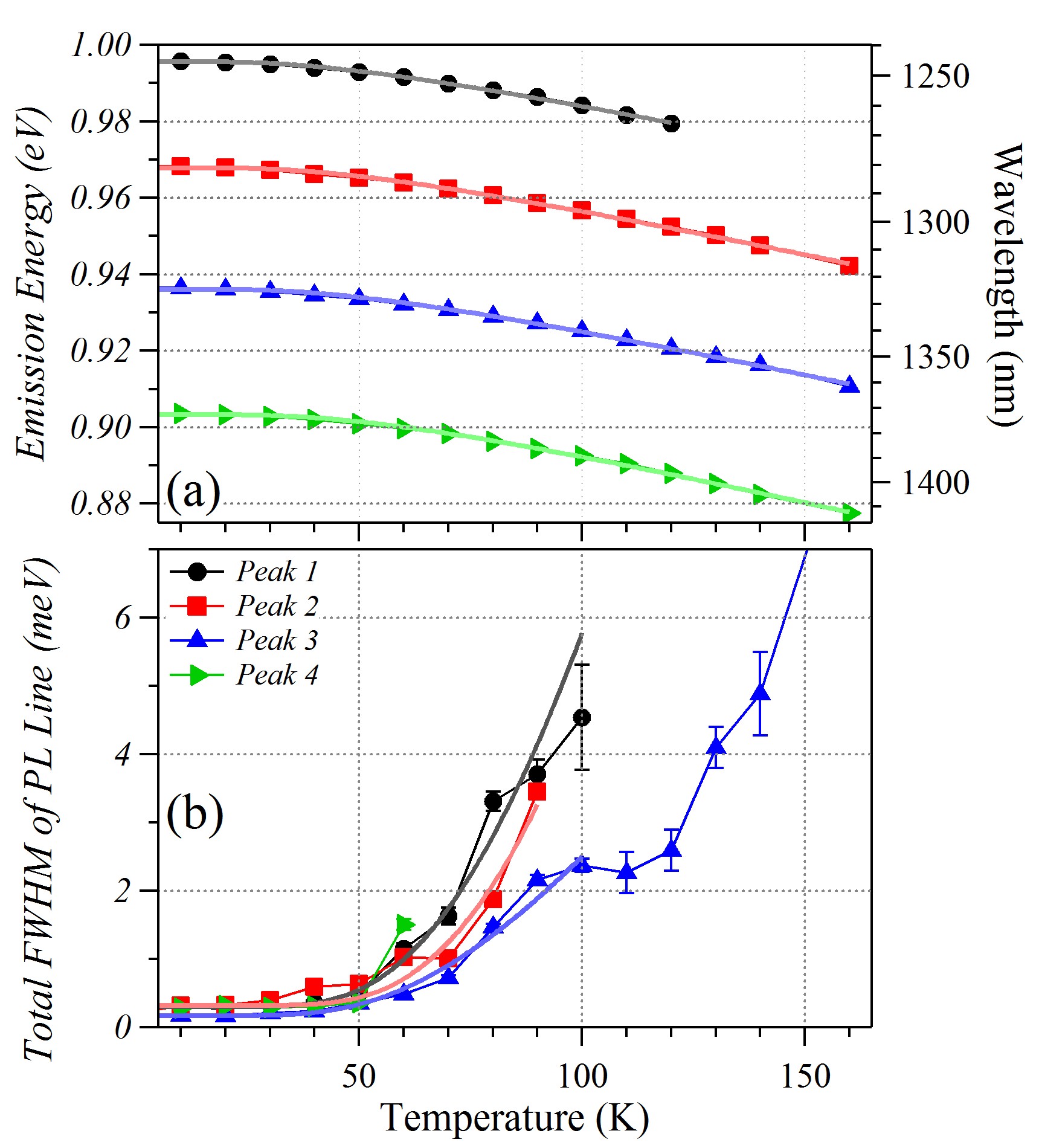}
	\caption{(a) The temperature dependent line shifts of four selected peaks, \emph{``Peaks 1--4''} as identified in \fref{fig:AaT001All}. (b) The line widths (FWHM) as a function of temperature for the same four selected peaks. All emission energy and line width values of the analyzed peaks are obtained from the data displayed in \fref{fig:AaT001All}. The corresponding fits to the PL energy and FWHM data, according to equations~(\ref{Eq:shift1}, \ref{Eq:fwhm1}) are shown as solid lines. Mind that the data points are only displayed in the temperature range up to the quenching of the respective peak and, for the FWHM data, until no more reliable evaluation due to peak overlap was possible.}
	\label{fig:AbLines}
\end{figure}

In the following paragraph, the temperature dependence of individual QD emission, exemplified by the four PL lines highlighted as \emph{``Peaks 1--4''} in \fref{fig:AaT001All}, is analyzed using phonon-carrier coupling. The respective emission energy and line width data as well as the corresponding fits are summarized in \fref{fig:AbLines}.
First, we discuss the change in the QD PL emission energies with increasing temperature (\fref{fig:AbLines} (a)).
Assuming phonon-carrier coupling as the most relevant mechanism, leads to an empirical fit function based on a Bose-Einstein statistical factor with an average phonon energy, ${<E_{\rm ph}^{\rm E}>}$, which causes the energy shift, as introduced, e.g., in \cite{Cin} (among various others), 
\begin{equation}
	E(T) = E_0 - \frac{S_{\rm c}}{\exp[{<E_{\rm ph}^{\rm E}>} / (k_{\rm B} T)] - 1}\ .
	\label{Eq:shift1}
\end{equation}
$E_0$ is the zero-temperature energy and $S_{\rm c}$ is the constant describing the strength of the electron-phonon coupling.
Since the fitted average phonon energy values, ${<E_{\rm ph}^{\rm E}>}$, of all peaks are between 12 and 14 meV, which is significantly lower than the bulk LO($\Gamma$) phonon energies of GaAs (35 meV) or InAs (29 meV), a significant contribution of acoustic phonons can be assumed (TA(X) of GaAs: 10 meV, InAs: 7 meV). The value of the coupling constant varies between $S_{\rm c}=$ 32 and 45 meV. The obtained result of a major contribution of the acoustic phonons is in good agreement with the afore discussed crossover between ZPL and accoustic phonon sideband dominated PL.

Second, we discuss the alteration and broadening of the QD emission line width (\fref{fig:AbLines} (b)).
The suggested model, again accounting for electron-phonon interaction mediated broadening, results in another empirical fit function \cite{Cin} similar to equation~\eref{Eq:shift1},
\begin{equation}
	\Gamma(T) = \Gamma_0 + \frac{\Gamma_1}{\exp[{<E_{\rm ph}^{\rm B}>} / (k_{\rm B} T)] - 1}\ ,
	\label{Eq:fwhm1}
\end{equation}
where $<E_{\rm ph}^{\rm B}>$ is the energy of the phonons causing the line broadening, $\Gamma_0$ is the zero-temperature line width (intrinsic broadening) and $\Gamma_1$ is the broadening constant. 

The fitting of \emph{``Peaks 1--3''} for temperatures up to around 100 K reveals similar values for the phonon energies, ${<E_{\rm ph}^{\rm B}>} = 22 - 32$ meV, with the average of around 27 meV, which coincides well with the bulk LO($\Gamma$) phonon energy of InAs, and which is also close to the bulk LO($\Gamma$) phonon energy of the InGaAlAs barrier \cite{Borr}. (Note that reliable line width values for \emph{``Peak 4''} could not be extracted for $T\geq 50$ K because of the increasing overlap with a neighbouring PL line, and that the respective data were thus not considered.)
The coupling constant, $\Gamma_1$, however, varies quite significantly between the peaks ($\Gamma_0 \approx 30 - 160$ meV) as it strongly depends on the exact environment of the respective dot, its size, potential, strain (anisotropy), local piezoelectric field, presence of defects, etc.
As can be seen in \fref{fig:AbLines}~(b), the line width of the peaks remains almost unchanged on the order of $0.1-0.3$ meV for temperatures below around $40$ K and then eventually increases more significantly above around $50$ K. 

Although such a simple discussion and fitting already gives a rather consistent description of the temperature dependence of the individual QD excitonic emission lines, it is not in complete agreement with the detailed discussion carried out above for the single recombination line, \emph{``Peak 3''}, and only provides supplementary information. It mainly describes the LO-phonon mediated pure dephasing process by inelastic electron-phonon scattering and thus the broadening of the ZPL, which is well supported by the obtained average phonon energy of 27 meV \cite{Borr,Sang2}. However, it does not account for the acoustic phonon sidebands and other broadening mechanisms and the respective line shape transition, because the ZPL is the dominant contribution in the main part of the temperature range investigated here. 
It shall also be added that it was rather difficult for this reason to clearly obtain and assign appropriate single values for all the line widths for temperatures above around 50 K, which is reflected by the uncertainties and deviation between data points and fit curves in \fref{fig:AbLines}~(b). 
Besides the afore discussed reasons, the increased broadening and intensity reduction for temperatures above $80-100$ K, which can be seen in figures \ref{fig:AbPeak3} and \ref{fig:AbLines}, can be due to charge fluctuations and an increased probability of various multi-phonon processes, that might meet resonance conditions between QD ground and excited states or the CES, which leads to additional non-radiative decay processes of the exciton states under investigation.

\subsubsection{Relative quantum dot intensities, coupling and quenching.}
\label{sec:RelativeQDIntensitiesCouplingAndQuenching}

\begin{figure}
	\centering
	\includegraphics[scale=.5]{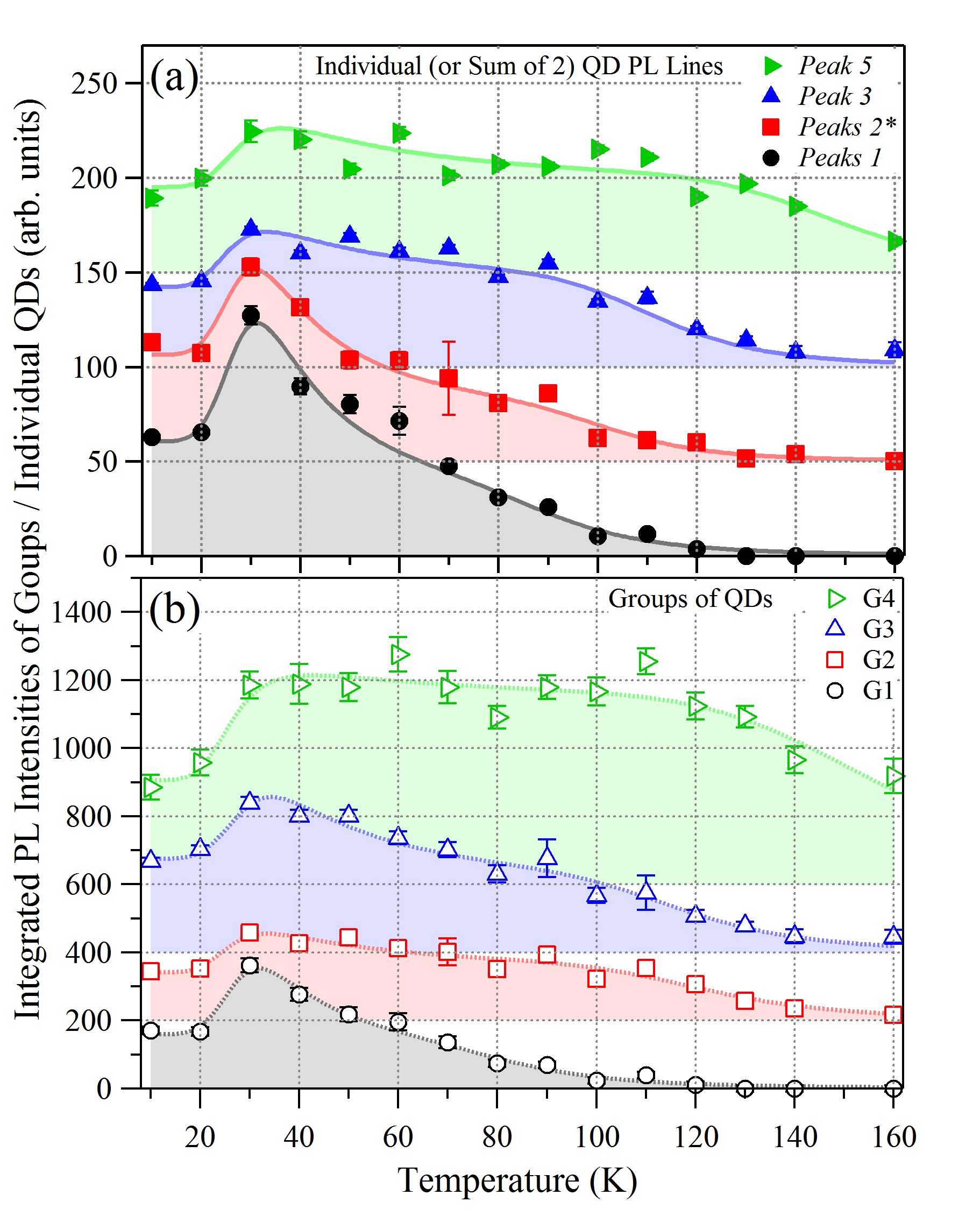}
	\caption{The PL intensity temperature dependence of (a) four individual QD peaks (or pairs) at different spectral locations, and (b) the integration over the QD groups 1--4, respectively; all intensities of the analyzed peaks and groups are obtained from the data displayed in \fref{fig:AaT001All}. The data sets are displayed as markers (with error bars) and are fitted using equation \eref{Eq:Arrha2}, the fit results are displayed as filled curves.  All data sets and corresponding fits are offset for clarity.}
	\label{fig:AcArrh}
\end{figure}

\Fref{fig:AcArrh} highlights the temperature dependence of the individual QD PL intensities under low power excitation, where excited states and re-excitation processes play no major role. It displays the experimental data for individual peaks (a) and dot groups (b) (as identified in \fref{fig:AaT001All}), as well as, the fits to all data sets using the afore introduced equation \eref{Eq:Arrha2}, which accounts for the initial QD carrier capture and for two distinct decay processes.
As a common trend of all displayed data sets, it can be observed that the lower energetic PL features persist visible in the spectra at increasing temperatures while the higher energetic features already start to fade.
The most significant fit parameters are summarized in table~\ref{tab:indivPL}.

As already introduced before, it is assumed that the excitations (excitons/charge-carriers) are transferred to the lower-energy dots as the temperature increases; this assumption is supported by the parameters corresponding to the first escape mechanism ($i=1$), e.g., the smaller activation energies, $E^{({\rm a}1)}$. All obtained values, for single dots and dot groups, range between 21 and 15 meV with a trend of the smaller values for the lower energy dots.
More notably, the corresponding coefficient, $b_1$, also gets smaller with decreasing PL energy, which indicates a smaller escape probability and larger (re-) capture-rate for the lowest energy dots and groups. This very well coincides with the carrier redistribution process from the high- to the low-energy dots as afore discussed and shown in \fref{fig:AXEns}~(b).

\begin{table}
	\centering \footnotesize
		\begin{tabular}{c|*{4}c|c} \hline 
			Type & $b_1$ & $E^{({\rm a}1)}$\footnotesize (meV) & $b_2$ & $E^{({\rm a}2)}$\footnotesize (meV) &  $\Delta E_{j}$\footnotesize (meV)\\ \hline
			\textit{Peak 5} & $602\pm8$ & $17.0\pm 0.7$ & $4.3\cdot10^7\pm0.5\cdot10^7$ & $162\pm18$ & \itshape$\approx$322\\
			\textit{Peak 3} & $985\pm32$ & $16.8\pm 0.2$ & $2.6\cdot10^7\pm0.2\cdot10^7$ & $117\pm20$ & \itshape$\approx$244\\
			\textit{Peaks 2$^{\ast\ \dagger}$} & $2683\pm254$ & $20.7\pm0.4$ & $6.3\cdot10^7\pm1.2\cdot10^7$ & $113\pm20$ & \itshape$\approx$218\\
			\textit{Peaks 1 $^{\dagger}$} & $3036\pm266$ & $21.2\pm 0.4$ & $9.0\cdot10^6\pm 1.0\cdot10^6 $ & $87\pm8$ & \itshape $\approx$184\\ \hline	
			Group 4 & $599\pm38$ & $15.5\pm0.4$ & $2.1\cdot10^6\pm0.6\cdot10^6$ & $128\pm32$ \\
			Group 3 & $809\pm79$ & $18.7\pm0.5$ & $8.4\cdot10^6\pm1.5\cdot10^6$ & $115\pm25$ \\
			Group 2 & $1010\pm115$ & $17.1\pm0.5$ & $7.9\cdot10^6\pm2.4\cdot10^6$ & $114\pm32$ \\ 
			Group 1 & $2665\pm555$ & $20.8\pm0.8$ & $3.0\cdot10^6\pm0.7\cdot10^6$ & $75\pm13$ \\ \hline 
		\end{tabular}
	\caption{Fit parameters (equation \eref{Eq:Arrha2}) for the individual QD PL lines and the energetically sorted groups of QDs; the $1\sigma$-fit-errors are omitted when of no significant magnitude ($<<1\%$ of the parameter value). The complete set of all fit parameters is displayed in~\cite{xtab}. The right column contains the optical gap energies between the InGaAlAs barrier and the respective QD PL peak. 
	
	$^{\dagger}$ The sum of the intensities of two neighbouring peaks is considered because of their spectral overlap at higher temperatures.}
	\label{tab:indivPL}
\end{table}

The coefficients corresponding to the second escape mechanism (and larger activation energies), $b_2$, which will be discussed in more detail below, are all rather large with no clear difference between dots or groups, i.e., all such escape processes happen with much higher probabilities and are dominated by the escape -- re-capture is much less likely for this process. 

\begin{figure}
	\centering
		\includegraphics[scale=.25]{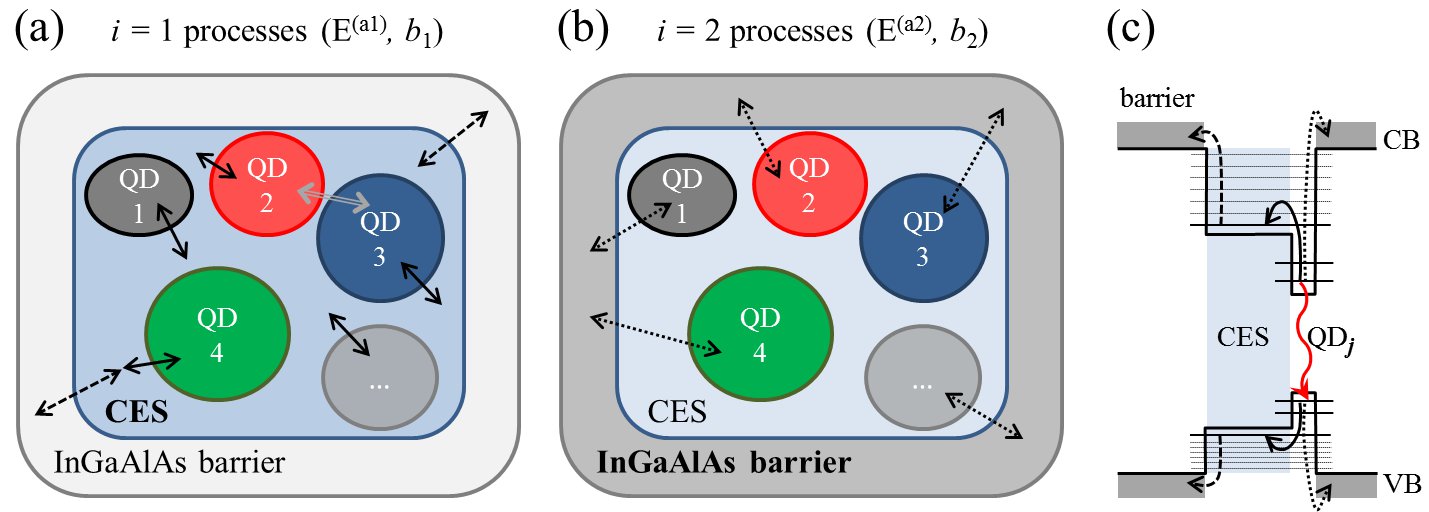}
	\caption{Schematic illustrations of the InGaAlAs barrier, the CES and some QDs inside a nano-cone, where (a) highlights the couplings related to the first carrier escape process ($i=1$) and (b) the couplings related to the second carrier escape process ($i=2$). (c) Illustration of a schematic simplification of the band diagram showing the assumed carrier escape, relaxation and coupling processes as well as energies for one QD and the common features CES and barrier.}
	\label{fig:AdScheme}
\end{figure}

The schematic \fref{fig:AdScheme} shall illustrate the different coupling mechanisms between dots, CES and barrier, and shall also indicate the corresponding confined discrete and continuous states as well as the assigned carrier transfer processes. The different arrows illustrate the various coupling mechanisms: 
1.) the inter-dot direct tunneling (grey double line arrow), 
2.) the inter-dot coupling via the CES (solid line arrows), 
3.) the coupling between CES and barrier (dashed line arrows) and 4.) the coupling between dots and barrier (dotted line arrows). 
1.) is not analyzed in this work and not furthermore considered because it is a local coupling effect between two distinct dots, which could be consequently treated as a new single quantum system, a lateral QD molecule \cite{q4,q5,q6}. 
2.) and 3.) represent the inter-dot carrier exchange via the CES and the possibility of subsequent carrier escape to the barrier (the first, $i=1$, escape process, \fref{fig:AdScheme} (a)), e.g., via phonon assisted ``shake-up'' of the carriers along the CES higher state density, as illustrated in \fref{fig:AdScheme}~(c). 
Finally, 4.) represents the carrier escape from the QDs to the barrier continuum (the second, $i=2$, escape process, \fref{fig:AdScheme} (b)).

The parameters assigned to the second escape mechanism ($i=2$), especially the larger activation energies, $E^{({\rm a}2)}$, of the individual dots (\fref{fig:AcArrh}~(a)) and groups of dots (b) clearly reveal this trend of decreasing activation energy with increasing QD emission energy.
This trend of the fitted activation energies is comparable to the trend of the respective optical energy  gaps, i.e., the energy separations between the dot PL energies and the barrier PL energy, $\Delta E_{j}$, which are listed in the right column in table \ref{tab:indivPL}. 
The generally larger relative $1\sigma$-fit-errors obtained for the dot groups reflect the fact, that the considered data contains PL from a certain number of individual QDs that only have the similar emission energies in common. Therefore, the fits have to account for some sort of ``averaging'' of a few dots with each very similar, but nevertheless slightly different values, especially with regard to the second escape mechanism.

In the following paragraph the values obtained for this second escape process, especially the activation energies, $E^{({\rm a}2)}$, are quantitatively analyzed and compared to the corresponding optical gap energies, $\Delta E_{j}$.
The carrier escape mechanisms from QDs can be indirectly represented by the ratio
\begin{equation}
	 \nu_j = E_j^{({\rm a}2)} / \Delta E_{j}\ ,
	 \label{Eq:nu}
\end{equation}
where the index $j$ represents a certain QD or QD group and moreover a corresponding optical gap energy (any energy difference between the described QD $j$ and some connected higher-energy state, the InGaAlAs barrier in the present case). See a comprehensive overview in G\'{e}linas, et al. \cite{Gel}.

Represented by this ratio of activation energy and energy gap between confined QD states and a barrier or wetting layer, different carrier escape scenarios can be assumed and supported by experiments \cite{San,Sch}.
In the case of (complete) exciton escape, the activation energy equals the full optical gap, i.e., $\nu_j = 1$. 
When in the actual escape process single carrier escape dominates, the activation energy $E_j^{({\rm a}2)}$ corresponds to the energy of the less confined carrier, i.e., $\nu_j < 1/2$.
Another escape mechanism, the correlated electron-hole pair escape, has been suggested with  $\nu_j = 1/2$. In this case, electrons and holes are assumed to be on average emitted as pairs, with their concentrations being equal or at least unchanged both within the dots and the barrier. This process leads to an activation energy of half the optical gap \cite{Mic2,Yan}.

In table \ref{tab:indivPL} we display the fitted activation energies, $E_j^{({\rm a}2)}$, besides the corresponding optical gap energies, $\Delta E_{j}$, which were obtained based on the low temperature QD peak PL and the barrier PL shown in \fref{fig:AaPowerAll}. Apparently, there exists a direct (linear) relation between these quantities; the fitted activation energies, $E_j^{({\rm a}2)}$, coincide well with  $\Delta E_{j}/2$. This correspond to $\nu_j = 1/2$ and thus clearly indicates a correlated electron-hole pair escape mechanism \cite{xx,Mic2,Yan}. However, since $E_j^{({\rm a}2)}$ is always either $\Delta E_{j}/2$ or slightly on the smaller side, single carrier escape of the less confined particles, most likely holes, might also be a reasonable scenario.

\section{Extension of quantum dot luminescence to around 1.55 $\mu$m}
\label{sec:1.55}

\begin{figure}
	\centering
		\includegraphics[scale=.5]{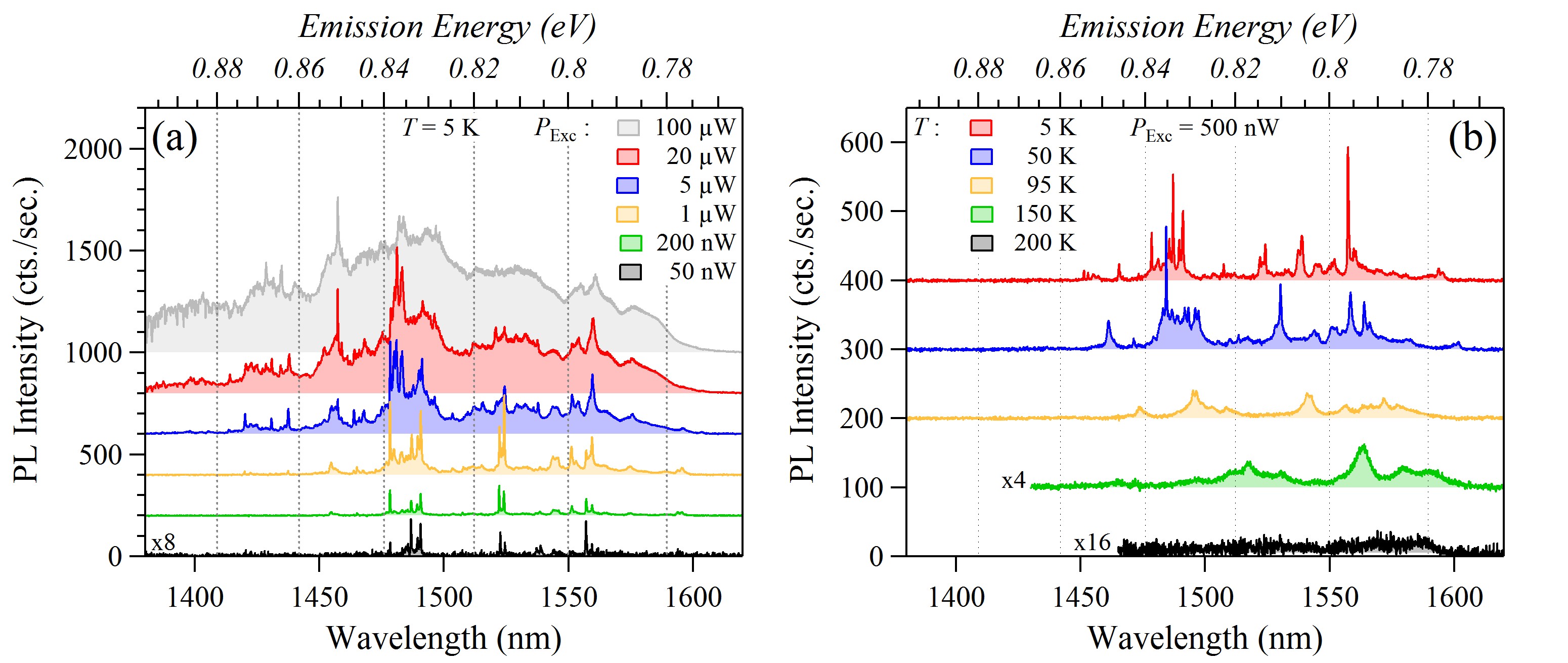}
	\caption{PL from 6 ML QDs at around 1.55 $\mu$m embedded in a $d_{\rm QD}\approx120$~nm  Si$_3$N$_4$/Ag-embedded nano-cone under non-resonant excitation. (a) Excitation power dependence at 5 K and (b) temperature dependence at 500 nW low power excitation.}
	\label{fig:6ML}
\end{figure}

Finally, we want to highlight the possibility that, using the same QD growth method, it is possible to not only cover photon wavelengths within the telecommunication O band, but to reach the technologically more interesting, however more challenging, C band around 1.55 $\mu$m (0.8 eV). This is realized by increasing the nominal deposition of InAs for the QD layer from four to six MLs. 
\Fref{fig:6ML} briefly summarizes the excitation power and temperature dependence from these types of lower energy QDs. Notably, QD emission persists up to around 150 K (or even 200 K). However, at the higher temperatures and thus longer PL wavelengths, the photon detection is limited by the cut-off of the used InGaAs detector, which is around 1.6 $\mu$m.

In comparison to the detailed presentation of the 1.3 $\mu$m QDs, a similar behavior for both the excitation and temperature dependence is found. The trend, however, suggests that the longer wavelength 1.55 $\mu$m QD PL is indeed even more temperature stable, i.e., the corresponding activation energies might be larger. Similar inter-dot coupling as demonstrated for the 1.3 $\mu$m QDs and therefore also similar escape mechanisms are expected.
A more detailed examination of the optical properties of such single QDs is currently in progress \cite{xx6MLSQD}.

\section{Summary and outlook}
\label{sec:Summary}

We have presented the fabrication and usage of metal-embedded nano-cones, in order to select a sub-ensemble of high-density InAs QDs emitting in the telecommunication O and C bands. Accessing such a reduced number of QDs, their electronic and optical properties could be examined in detail. Individual dot emission has been demonstrated between 1.2 and 1.6 $\mu$m ($1.03-0.77$ eV). The QD PL persists up to temperatures of around 150 K; the transition of the spectral properties of such individual PL lines has been analyzed, accounting for the alteration of the transition energies, the line widths and shapes. The transition of a single QD PL line from the ZPL to an acoustic sideband dominated line shape could be demonstrated. 

The PL intensity temperature dependence of both the sub-ensemble of dots in a nano-cone and individual QDs has been investigated and could be fitted with a model that includes the carrier capture and two different carrier escape processes. One of the observed escape processes could be closely connected to the inter-dot coupling of excited electronic states, which not only leads to a carrier redistribution towards the lower energy dots but also contributes to the non-radiative carrier escape. These mechanisms are characteristic for samples with high spatial dot density, which enhances the lateral inter-dot coupling. The other observed carrier escape process is suggested to be due to correlated electron-hole pair escape and coincides well with half the optical gap energy between the InGaAlAs barrier and the respective QDs.

With this more detailed understanding on the QDs electronic properties we believe that the presented QDs, incorporated in nanometer-sized metal-embedded mesas, provide a suitable future realization for QD-based single photon sources in the telecommunication bands. Moreover, we have shown that they have the potential for application at elevated temperatures, which well covers the liquid Nitrogen temperature of 77 K, and that temperature tuning can be used to adjust the emission energy of the dots by at least ten meV. Especially the dots on the low-energy side of the sub-ensemble, which is usually the one of particular interest and importance, provide the best performance with regard to temperature stability. 
In the future we are planning to proceed towards the realization of single photon and entangled photon pair sources for application in quantum information processing and quantum communication using silica based fibre networks.

\ack 
This work was partially supported by SCOPE 
from the Ministry of Internal Affairs and Communications, 
by the Grand-in-Aid for Scientific Research (A), No.21246048, and 
by HINTS from the Ministry of Education, Culture, Sports, Science and Technology (MEXT), Japan.
CH acknowledges the Japan Society for the Promotion of Science (JSPS) for providing financial support in the form of a JSPS Fellowship for Foreign Researchers; NAJ acknowledges financial support via a MEXT scholarship.


\begin{thebibliography}{11}
\bibitem{xx6MLSQD} Our actual work is focusing on improving the long-wavelength samples to obtain efficient single QD PL extraction around 1.55 $\mu$m for usage as single photon/entangled photon pair sources.\\
\end{thebibliography}
\end{document}